**Biographies of Albert Einstein – Mastermind of Theoretical Physics**

Galina Weinstein

Over the years many have written biographies of Einstein. They all based their biographies on primary sources, archival material: memories and letters of people who were in contact with Einstein, Einstein's own recollections; interviews that Einstein had given over the years, and letters of Einstein to his friends – youth friends like Marcel Grossman and Michele Besso and later friends and colleagues like Heinrich Zangger; and especially his love letters with Mileva Marić. One can demarcate between two types of biographies, namely, Documentary biographies, and, Non-documentary biographies. Non-documentary biographies were written by people who based themselves on documentary biographies and on other non-documentary biographies. Documentary biographies were written by people who knew Einstein personally, and received information from him and from other people who were in personal contact with him. This type can be further divided into two subgroups: books that were written while Einstein was still alive, and books that were written after his death. Among the first group there were some biographies that Einstein considered to be more reliable, others he thought were less reliable, and still other biographies which he considered to be almost pure imagination. I will demonstrate that relying on primary sources and "documentary" does not necessarily mean "reliable".

Einstein wrote on February 22, 1949 to the writer Max Brod, "There have already been published by the bucketful such brazen lies and utter fictions about me that I would long since have gone to my grave if I had let myself pay attention to them".[1]

I will first give two examples of non-documentary biographies, and then address the question of *reliability* of documentary sources.[2]

## 1. Isaacson's Einstein

The first example of a non-documentary biography is **Walter Isaacson**'s latest biography *Einstein, His Life and Universe*. Isaacson has done a wonderful job of collecting as many archival documents as possible. He also spoke with major Einstein scholars, but wrote for instance, as to Einstein's spouse's Mileva Marić's role in Einstein's work, "Einstein happily praised his wife's help."I needed my wife", he told her friends in Serbia. "She solves all the mathematical problems for me".[3] And the source was **Dennis Overbye**'s historical romance, *Einstein in Love*, in which Overbye wrote, "Not long ago we finished a very significant work that will make my husband world famous", Mileva told her father in a conversation widely repeated through the years. […] Now she had brought back a handsome, adoring husband. Albert knew how to play the crowd. "I need my wife", he is reported to have said, "She solves all the mathematical problems for me";[4] and so they lived happily ever after in the world of the historical *romance*; and unfortunately, in some papers that quoted this last sentence. Overbye quotes a dubious source, *Im Schatten Albert Einsteins: Das tragische Leben der Mileva Einstein-Marić* (*In the Shadow of Albert Einstein: The Tragic Life of Mileva Marić*), which was translated to English for him.

---

[1] Dukas Helen, and Hoffmann, Banesh, Albert Einstein, *The Human Side, New Glimpses from his Archives*, Princeton: Princeton university Press, p. 21; p. 126.
[2] Reprinted in Hebrew abridged form in http://odyssey.org.il/?issue=224231: Weinstein, Gali, *Odyssey*, Vol. 14, January, 2012.
[3] Isaacson, Walter, *Einstein His Life and Universe*, 2007, New York: Simon and Schuster, p. 136.

**2. Fölsing's Einstein**

Isaacson frequently cited another important non-documentary biography, **Albrecht Fölsing**'s *Albert Einstein, A Biography*, 1997.[5] This is an abridged translation of Albrecht Fölsing, *Albert Einstein, Eine Biographie*, 1993.[6]

However, Fölsing based himself on non-documentary biographies too, such as **Max Flückiger**'s *Albert Einstein in Bern*.[7] Abraham Pais described Flückiger's book, "Contains a number of reproductions of rare documents pertaining to Einstein's younger days. The text contains numerous inaccuracies".[8]

How did Flückiger find these reproductions? He participated in the relativity congress in Bern that took place on July 11, 1955, celebrating 50 years of the special theory of relativity. He writes about it in his book and gives a list of participants.[9] Flückiger included a photo of himself in the conference.[10]

Flückiger probably spoke with some of these participants during the congress, and perhaps they told him some stories, which he could have integrated into his book.

Towards the end of Flückiger's book, he reproduced the reminiscences of people who knew Einstein personally. Of special interest is a radio broadcast by **Joseph Sauter**, a Swiss of French nationality, who worked with Einstein at the Patent office. Flückiger took the recording and made a transcription: "Comment j'ai appris à connaître Einstein".[11] Flückiger included a photo of the elderly Sauter (age 84) at the 1955 relativity congress, the same year that Sauter gave the radio broadcast.[12]

In his biography Fölsing misinterpreted Sauter's memoirs, and thus had added details which Sauter had never said.

---

[4] Overbye, Dennis, *Einstein in Love*, 2000, New-York: Viking Press, p. 140; Overbye refers to a source from 1993.
[5] Fölsing, Albrecht, English abridged translation by Ewald Osers: *Albert Einstein, A Biography*, 1997, New York: Penguin books.
[6] Fölsing, Albrecht, *Eine Biographie Albert Einstein*, 1993, Frankfurt am Main: Suhrkamp.
[7] Flückiger, Max (1960/1974), *Albert Einstein in Bern*, 1974, (Switzerland: Verlag Paul Haupt Bern).
[8] Pais, Abraham, *Subtle is the Lord. The Science and Life of Albert Einstein*, 1982, Oxford: Oxford University Press, p. 49.
[9] Flückiger, 1960/1974, pp. 182-183. Among the participants were known scientists and colleagues who were in contact with Einstein: Niels Bohr, Herman Bondi, Emile Borel, Max Born, Adrian Fokker, Paul Gruner, Leopold Infeld, Louis Kollros, Joseph Laub, Max von Laue, Wolfgang Pauli, Nathan Rosen, Joseph Sauter, Herman Weyl, Eugene Wigner, and others.
[10] Flückiger, 1960/1974, p. 185.
[11] Sauter, Joseph, "Comment j'ai appris à connaître Einstein", 1960, in Flückiger, 1960/1974, p. 153.
[12] Flückiger, 1960/1974, p. 155.

Another good example of how non-documentary biographies cite other non-documentary biographies is the following. Fölsing cited Constance Reid, a popular American biographer who wrote that Herman Minkowski, (one of Einstein's mathematics professors at the Zürich Polytechnic who later re-casted special relativity in a spacetime formalism) later often told his students in Göttingen, "Einstein's presentation of his deep theory [of relativity] is mathematically awkward – I can say that because he got his mathematical education in Zürich from me".[13] Reid did not give reference to this quotation and did not say from whom she heard this. It was probably just another of the "Einstein anecdotes" widespread in Göttingen. Reid wrote her book in English, and she visited in Göttingen; suppose that she heard this anecdote in Göttingen, then she had somehow to translate the quotation from German to English.

In 1993 Fölsing wrote his Einstein biography in German. He thus had to translate the Reid quotation into German: "So belehrte Hermann Minkowski seine Studenten in Göttingen: 'Einsteins Darstellung seiner tiefsinnigen Theorie ist mathematisch umständlich – ich darf das sagen, weil er seine mathematische Ausbildung bei mir in Zürich erhalten hat'".[14] Fölsing's German biography was then translated in 1997 to English. The Reid quotation was then translated from German back to English. Here is the 1997 result: "Thus Hermann Minkowski told his students in Göttingen: 'Einstein's presentation of his subtle theory is mathematically cumbersome – I am allowed to say so because he learned his mathematics from me in Zürich'.[15]

What then did Minkowski tell his students? Did he actually tell them something about Einstein?...

**3. Moszkowski's Einstein**

The earliest biographical sketch was written in 1921. A journalist **Alexander Moszkowski** had conversations with Einstein in Berlin between 1919 and 1920,[16] and entitled the sketch, "Der Werdegang und die Persönlichkeit" (The Career and Personality).[17] This sketch appeared in a book with the long title, *Einstein, Einblicke in seine Gedankenwelt. Gemeinverständliche Betrachtungen über die Relativitätstheorie und ein neues Weltsystem. Entwickelt aus Gesprächen mit Einstein*, or in short, *Gesprächen mit Einstein* [*Converstaions with Einstein*].[18]

Moszkowski's biographical sketch is important because it was written only a few years after the events in question.

---

[13] Reid, Constance, *Hilbert*, 1970/1996, New-York: Springer-Verlag, p. 112.
[14] Fölsing, 1993, p. 353.
[15] Fölsing, 1997, p. 311.
[16] Moszkowski wrote, "The discussions began in the summer of 1919 and ended in the fall of 1920". Moszkowski, Alexander (1921a), *Einstein, Einblicke in seine Gedankenwelt. Gemeinverständliche Betrachtungen über die Relativitätstheorie und ein neues Weltsystem. Entwickelt aus Gesprächen mit Einstein*, 1921, Hamburg: Hoffmann und Campe/ Berlin: F. Fontane & Co, p. 240.
[17] Moszkowski, 1921a, pp. 218-230.
[18] Moszkowski, 1921a.

Moszkowski was a journalist who, during World War I invited Einstein to join a literary circle in Berlin: "Einstein was appointed professor of the Academy of Sciences with the right of lecturing at the University of Berlin. This brought my personal wish within reach. Trusting to good fortune, I set about materializing it. In conjunction with a colleague I wrote him a letter asking him to honor with his presence one of the informal evenings instituted by our Literary Society at the Hôtel Bristol. Here he was my neighbor at table, and chatted with me for some hours."[19] Later, Moszkowski wrote articles about the solar eclipse and the expedition for a major newspaper the *Berliner Tageblatt*.

Moszkowski had published books on jokes and occult matters. When Einstein's friends heard the announcement of the forthcoming publication of a book by Moszkowski on Einstein, which was based on his conversations with Einstein during the war, his friends were horrified. Einstein himself was away on one of his trips. Einstein's close friend Max Born, and especially his wife Hedi Born, tried to persuade him to prohibit publication of the book, as Moszkowski was an unpleasant author that recorded his conversations with Einstein in a book without the latter's authorization: "You must withdraw your permission to X for the publication of the book *Conversations with Einstein*, and what's more *at once*, and by *registered* letter".[20] The book will thus provide the best confirmation of the accusations (on behalf of Phillip Lenard, Paul Wyland, Ernst Gehrcke, and other Anti-Semites) Einstein was trying to put himself up. Einstein thought Hedi Born's verdict on Moszkowski was too harsh, but he obeyed, and sent him a registered letter, asking him not to print his book.[21]

Neither Moszkowski nor his publisher agreed to withdraw the publication. Einstein was advised to take legal action, but he rejected that idea so that it would not magnify the scandal: "[Paul] Ehrenfest and [Hendrik Antoon] Lorentz advise against legal proceedings". After all Einstein thought "nothing can happen to me. In any case, I have used the strongest means at my disposal, apart from legal ones, particularly the threat that I would break off our relationship. However I still prefer X to Lenard and Wien […] the former does it only to earn money (which is, after all, better and more reasonable)".[22] When Max Born finally picked up the book in old age he realized that it was not bad as he expected. After all Moszkowski learned many of the details from conversations with Einstein.[23]

Moszkowski's book was immediately translated to English by Henry L. Brose in the same year 1921 under the title, *Einstein the Searcher His Works Explained from Dialogues with Einstein*, and published in London.

---

[19] Moszkowski, Alexander (1921b), *Einstein the Searcher His Works Explained from Dialogues with Einstein*, 1921, translated by Henry L. Brose, London: Methuen & Go. LTD; appeared in 1970 as: *Conversations with Einstein*, London: Sidgwick & Jackson, 1970, p. 2; Moszkowski, 1921a, p. 16.
[20] Hedi Born to Einstein, October 7, 1920, Einstein, Albert, *Albert Einstein Max Born Brief Wechsel 1916-1955*, 1969/1991, Austria: Nymphenburger, letter 23, pp. 61-63.
[21] Einstein to Max Born, October 11, 1920, Einstein, 1969/1991, letter 25, p. 65.
[22] Einstein to Max Born, Einstein, 1969/1991, undated, letter 26, p. 65.
[23] For details of the story see Fölsing, 1997, pp. 469-471

Brose wrote in the introduction,[24]

"[…] there is even now a considerable amount of literature about him [Einstein]. At the end of this generation we shall possess a voluminous library composed entirely of books about Einstein. The present book will differ from most of these, in that Einstein here occurs not only objectively but also subjectively. We shall, of course, speak of him here too, but we shall also hear him speak himself, and there can be no doubt that all who are devoted to the world thought can but gain by listening to him.

The title agrees with the circumstance to which this book owes its birth. And in undertaking to address itself to the circle of readers as to an audience, it promises much eloquence that came from Einstein's own lips, during hours of social intercourse, far removed from academic purposes and not based on any definite scheme intended for instruction. It will, therefore, be neither a course of lectures nor anything similar aiming at a systematic order and development. Nor is it a mere phonographic record, for this is made impossible if for no other reason than that whoever has the good fortune to converse with this man, finds every minute far too precious to waste it in snatching moments to take shorthand notes. What he has heard and discussed crystallizes itself in subsequent notes, and to some extent he relies on his memory, which would have to be extraordinarily lax if it managed to forget the essentials of such conversations".

**4. Maja's Biography of her Brother**

Einstein's sister, **Maja Winteler-Einstein**, wrote a biographical sketch (*Albert Einstein – beitrag für sein lebensbild*, *Albert Einstein – his contribution to the Picture of his life*) in 1924.[25] She completed her draft on February 15, 1924, still unfinished. It is the earliest known draft of a planned biography and covers Einstein's life until 1924. In the foreword Maja notes that she placed the main emphasis on the description of Einstein's youth and the milieu from which he came. As Abraham Pais wrote, "Maja's biographical essay about her brother, completed in 1924, is the main source of family recollections about Albert's earliest years".[26]

The first 17 pages of the biography are included in the first volume of the *Collected Papers of Albert Einstein* (*CPAE*).[27] The complete original is kept by the "Besso estate" (not reachable and hard to find), but a typed copy is in the Einstein Archives. The copy is not in a good physical condition. A few pages are missing, few are duplicated, and the typing is sometimes blurred.

---

[24] Henry L. Brose, in Moszkowski, 1921b, p. vi.
[25] Winteler-Einstein Maja, *Albert Einstein –Beitrag für sein Lebensbild*, 1924, Einstein Archives, Jerusalem: the full biography printed in a typing machine with a few missing pages and double pages.
[26] Pais, 1982, p. 36.
[27] Winteler-Einstein Maja, *Albert Einstein –Beitrag für sein Lebensbild*, 1924, reprinted in abridged form in *The Collected Papers of Albert Einstein Vol. 1: The Early Years,1879–1902* (*CPAE*, Vol. 1)*, Stachel, John, Cassidy, David C., and Schulmann, Robert (eds.), Princeton: Princeton University Press, 1987, pp xlviii-lxvi.

One main characteristic of Maja's biography is her very emotional involvement. Maja was very involved when her brother could not find a position after he graduated. For instance, she wrote that "his nutrition was very poor", and "then he started with his stomach illness".[28] And, after her brother's relativity paper was published in 1905, Maja said, that, scientists sent their letters to "Professor Einstein" and directed them to the University of Bern, "because no one suspected that the author of the publications that by then had aroused great stir was in a humble official position in the Patent Office".[29] Maja thus cared about her brother who was poor, got stomach illness and sat in a humble office, while he was so gifted; she felt that scientists around did not fully appreciate her brother at that time. In addition, she left some things she might have known such as the child out of wedlock – Lieserl.

**5. Reiser's Einstein**

The journalist **Rodulf Kayser**, the husband of Einstein's stepdaughter Ilse (writing under the pseudonym **Anton Reiser**), wrote a biography with Einstein's approval and his so-called cooperation.[30] However, Einstein did not consider Reiser's biography as completely reliable. Einstein gave the book his qualified endorsement: "The author of this book is one who knows me rather intimately in my endeavor, thoughts, beliefs – in bedroom slippers". However, Einstein also wrote in the preface to Reiser's book: "I found the facts of this book duly accurate, and its characterization, throughout, as good as might be expected of one who is perforce himself, and who can no more be another than I can".[31]

Einstein refused to let Reiser publish his biography in German, and for this reason the biography was translated into English by the American poet Louis Zukofsky, and first published in America in 1930, and then in London in 1931.

Einstein wrote on May 2, 1932 on Reiser's book that he did not accuse the book as "Geschmacklosigkeit", bad taste, but,[32]

"Generally I find it tasteless, when biographies of or autobiographies by still living persons are published. This is except for those cases where it is a matter of the representation of events or relations, in which the personal is put in the background"

Einstein did not like biographies that dealt too much with his personal life. He then said,[33]

---

[28] Winteler-Einstein, 1924, p. 20.
[29] Winteler-Einstein, 1924, pp. 23-24.
[30] Reiser, Anton *Albert Einstein: A Biographical Portrait*, 1930/1952, New York: Dover.
[31] Reiser, 1930/1952, Einstein's preface.
[32] "Allgemein aber empfinde ich es als geschmacklos, wenn über noch lebende Biographisces oder Autobiographisches publiziert wird. Ausgenommen sind solche Fälle, wo es sich um Darstellung von Begeben heiten oder irgend welchen verhältnissen handelt, die das Persönliche in den Hintergrund treten lassen". Reiser, Anton, *Albert Einstein, Ein Biographisches Porträt*, 1997, New-York: Albert & Charles Boni, unpublished manuscript, pp. 226-227.

"I have also prohibited Reiser's book from appearing in the German language, but allowed the book to appear in foreign languages, I also hold the latter to be quite tasteless. [… Reiser] need[s] to earn money, which serves as an excuse for and for that […he] cannot wait until I'm dead. Is the mention of such a basic fact an accusation?"

In 1997 Rudolf Kayser's second wife, Eva Kayser (who died two years later in 1999), was asked for permission to publish the book in a German translation. Mrs. Kayser gave the following reply,[34]

"New-York, January 22, 1997

Dear Mr. Schwarz,

[…] Unfortunately it is not possible for me to authorize a publication of Rudolf Kayser's (Anton Reiser) 'Albert Einstein. A Biographical Portrait' in a German translation. I remember numerous conversations with my husband, from which it was clear that both he and Professor Einstein did not desire a German translation of the book, and wished to confine it to the English language."

A preprint in German of Reiser's book is in the Einstein Archives. The pre-print was supposed to be published in 1997 as a book by "Albert & Charles Boni" in New-York.[35] The German version has some differences from the English one; in particular there are many black and white photos.

## 6. Frank's Einstein

These biographies and contributions to Einstein's biography are the earliest biographical descriptions of Einstein. They are important because they describe Einstein and his work in close proximity to the events – within 15 to 20 years after Einstein's path breaking papers. A common characteristic of these biographies is that they were all written while Einstein was still in Berlin.

In 1933 Einstein emigrated to the United States. A few years later documentary biographies of Einstein began to be published. The first important biography was written by **Philip Frank**, a physicist-philosopher from the former Vienna circle who wrote the biography after he emigrated to the United States.[36]

---

[33] "Ich habe auch verboten, daß das Reisers Buch in deutscher Sprache erscheint und andererseits Ihnen erlaubt, Ihr Buch in fremden Sprachen erscheinen zu lassen. Letzteres halte ich Zwar auch für geschmak los. In beiden Fällen dient ober als Entschuldigung, dass die Autoren es wirklich nötig hatten bzw. haben Geld zu verdienen, und dass sie damit nicht warten könen, bis ich tot bin. Ist die Erwähnung einer so elementaren Tatsache ein Vorwurf?" Reiser, 1997, pp. 226-227.

[34] "Ich erinnere mich Zahlreicher Unterredungen mit meinem Mann aus denen klar hervorging, daß sowohl er sowie Professor Einstein eine Veröffent lichung des Buches in einer deutschen Übersetzung nicht wünschten und sich auf die englische Sprache beschränken wollten". Reiser, 1997, p. 232.

[35] Reiser, 1997.

[36] Frank, Philip, *Einstein: His Life and Times*, 1947, New York: Knopf, 2002, London: Jonathan, Cape.

In 1912 Frank became Einstein's successor as professor of theoretical physics at the University of Prague, where he stayed until 1938. When Frank came to the United States and met Einstein again, he then conceived the idea of taking advantage of the physical proximity to prepare an account of Einstein's life and work. "When I told Einstein about this plan he said: 'How strange that you are following in my footsteps a second time!' ".[37] Frank composed the biography in German. However, it first appeared in 1947 in an English translation that omitted parts of the German manuscript. Subsequently, the German complete manuscript was published in 1949 under the title, *Albert Einstein sein Leben und seine Zeit* with a forward by Einstein.[38]

Generally, Einstein disliked biographies. In the forward to Frank's biography he explained why he did not like biographies,[39]

"I must confess that biographies have rarely attracted or fascinated me. Autobiographies mostly owe their existence to the feeling of negative character against one's fellow humans or to self-love. Biographies from the pen of another person usually reflect in their psychological traits more the intellectual and moral habits of the writer than those of and the person described; the bridge of understanding between two individualities is usually weak […]".

However, Einstein said on Frank's biography,[40]

"What I believe, however, that I can promise the reader, is this: he will find sage, interesting and plausible explanations in this book, which will be at least in part new and surprising to him."

Einstein said that, "It is a biography of a person [Einstein], whose life portrays a chain of actions", and whose "company is not so bad". The reader will read about his "personal relationships with other human beings" and also with organizations.[41]

Einstein explained further a most important element for him in Frank's biography,[42]

---

[37] Frank, 1947/2002, pp. xiii-xiv.
[38] Frank, Philip, *Albert Einstein sein Leben und seine Zeit*, 1949/1979, Braunschweig: F. Vieweg.
[39] "Ich muß bekennen, daß Biographien mich selten angezogen oder gefesselt haben. Autobiographien verdanken ihre Entstehung meist der selbstliebe oder Gefühlen negativen Charakters gegen Mitmenschen. Biographien aus der Feder einer anderen Person spiegeln in ihren psychologischen Zügen meist mehr den intellektuellen und seelischen habitus des Schriftstellers als den der geschilderten Person; die Brücke des verstehens Zwischen zwei individualitäten ist meist schwach [...]". Einstein, vorwort, Frank, 1949/1979.
[40] "Was ich aber glaube, dem Leser versprechen zu dürfen, das ist dies: er wird kluge, interessante und plausible Erklärungen in diesem Buche finden, die ihm wenigstens zum Teil neu und überraschend sein warden". Einstein, vorwort, Frank, 1949/1979.
[41] "Handelt es sich um die biographie einer Persönlichkeit, deren Leben eine Kette von Handlungen darstellt, deren Wirkungen weithin sichtbar sind, so ist das Unternehmen nicht gar so schlimm; es ist ebenso in Fällen, in denen die persönlichen Beziehungen zu anderen Menschen und menschlichen Organisationen eine Hauptrolle spielen". Einstein, vorwort, Frank, 1949/1979.
[42] "Soll aber eine Person dargestellt werden, bei welcher die Bemühung um Probleme der Erkenntnis den wichtigsten Teil des Erlebens darstellen, dann ist das Unternehmen des Biographen wahrlich nicht beneidenswert und aussichtsreich". Einstein, vorwort, Frank, 1949/1979.

"But if a person is presented when the most important problems are dealt, the company of the biographer is certainly unnoticed and imperceptible."

Frank was not distracted or inhibited too much by unnecessary personal details. He revealed professional and human ties in Einstein's life, rather than personal stories. These according to Einstein meant a lot more to a biography than personal details. Frank was able to reveal the professional ties because he was an experienced physicist. Einstein said at the beginning of the Forward, he can promise the reader that, he would find interesting explanations in Franks' book, which will be at least in part new and surprising – and Einstein presumably meant interesting new and surprising explanations about the professional ties in his life.

Einstein goes on to say in the Forward to Frank's book that, he could not find a decent biography before that of Frank's, alluding probably to Reiser's biography. Einstein wrote, "I am inclined to answer this question in no, as far as my own person is concerned; so no biography!". [43]

Gerald Holton tells the story about his Harvard university mentor Philipp Frank, "his wife had suggested he write" the biography of Einstein. Holton probably heard this story from Frank himself.[44] However, in the forward to Frank's book Einstein claims that he encouraged Frank as to parts of this biography.[45] So who gave Frank the idea to write the biography? Very likely both Frank's wife and Einstein; Einstein was not a biography lover, but he preferred biographies explaining the history of his scientific ideas upon biographies explaining his personal life.

Holton writes, "That book is still one of the best, not least by virtue of giving the historical, philosophical and cultural context, even though the manuscript of his book was horribly mangled by its publisher in the English-language edition. Alfred A. Knopf [the publisher] gave the manuscript to edit to an American who, Philipp told me, knew English but no science, and also to a Japanese, who knew science but no English. Between the two, Philipp's original German manuscript was severely truncated. Happily, List Verlag (and later Vieweg) in Germany published Philipp's original manuscript of his book, and even was persuaded, in a recent reprinting, to include Einstein's Preface, which Knopf had vetoed for the American edition".[46]

---

[43] "Ich bin sehr geneigt, diese Frage mit Nein zu beantworten, soweit meine eigene person in Betracht kommt; also keine Biographie!". Einstein, vorwort, Frank, 1949/1979. In 1952, Einstein wrote Carl Seelig quite the same thing about Frank's biography (quoted in full further below), "Andere einigermassen verlässliche Biographien existieren überhaupt nicht" (other reliable biographies do not exist). Item 39 011, Einstein Archives.
[44] Holton, Gerald, "Philip Frank at Harvard University: His Work and his Influence", *Synthese* 153, 2006, pp. 297-311; p. 300.
[45] "Und trotzdem ermutigte ich meinen alten Kameraden Professor Frank, dieses buch zu scheiben, das sozusagen den Anschein einer Biographie hat". Einstein, vorwort, Frank, 1949/1979.
[46] Holton, 2006, pp. 302-303.

## 7. Seelig's Einstein

The second documentary biography written after Einstein's immigration to the United States was written by the Swiss journalist and writer from Zürich, **Carl Seelig**. Unlike Frank, Seelig was *not* a physicist, nor did he have close relation with Einstein. He had initially tried to contact Einstein immediately after he became world famous in 1919. Seelig wrote Einstein on December 21, 1919 a flattering letter. He wanted Einstein to publish a book with his general contributions: [47]

"My deer Professor"

Seelig told Einstein that, he learned from the papers that Einstein became closely attached with the Clarté movement of Henri Barbusse of which Seelig himself was also related. Seelig presented himself as an editor of a collection named "The Twelve Books". These books, said Seelig, are the contributions by the most prominent international minds. Einstein would contribute a work of his own choice and Seelig would be the editor. Seelig gave Einstein some examples of books he had edited. Subsequently, Seelig went straight to business; he offered Einstein: for every 1000 copies the highest honorarium: earning 2000 Marks. Seelig said that a book edited by Seelig's publication of Einstein's contributions would bear a photo and a (handwritten) dedication, and he wanted to send Einstein a sample book from this series, that of Barbusse.

On December 29, 1919, Einstein politely rejected Seelig's offer:[48]

Einstein tells Seelig that he welcomes the "Clarté" movement, and he will do his best to advertize it in Germany; but unfortunately he is so busy at work that he cannot even think of writing the book Seelig wishes, as much as he would like to.

Einstein responded to the sample book sent to him by Seelig. He told Seelig that Barbusse photography with handwritten dedication instilled in him great sympathy for this great man. Einstein told Seelig that the books kindly promised to him would please him exceedingly, and even more so he would be happy to know him more when the latter comes to Zurich.

Seelig continued to pursue Einstein and pleaded with him to publish a book in addition at least in two letters in 1923.[49] Einstein sent him a polite reply in a few lines, one of which says, "I thank you for the kind offer".[50]

---

[47] Item 39 002, Einstein Archives, Seelig to Einstein, December 21, 1919, *The Collected Papers of Albert Einstein. Vol. 9: The Berlin Years:Correspondence, January 1919–April 1920* (*CPAE*, Vol. 9), Buchwald, Diana Kormos, Schulmann, Robert, Illy, Jószef, Kennefick, Daniel J., and Sauer, Tilman (eds.), Princeton: Princeton University Press, 2004, Doc. 230.
[48] Item 39 003, Einstein Archives, Einstein to Seelig, December 19, 1919, *CPAE*, Vol. 9, Doc. 237.
[49] Items 39 005, 39 006, Einstein Archives.
[50] Items 39 007 (16.4.23), Einstein Archives.

In 1952 Seelig finally managed to persuade Einstein to publish his 1934 *Mein Weltbild* in extended edition. The 1934 book was a collection of general papers and talks by Einstein to which the latter added in 1952 new papers and talks.[51] The book was published by Seelig's "Europa Verlag" in Zurich with Seelig as editor.[52] In 1953 when the expanded version was published, Seelig sent Einstein the manuscript which he corrected before publication.[53]

In 1952 Seelig and Einstein exchanged letters about the question of *a biography*. At that time Seelig developed a personal relationship with Einstein and took care of his youngest ill son Eduard (or "Tete") institutionalized in Zurich.[54]

However, Einstein seemed reluctant to disclose details in the exchange with Seelig pertaining to his private life. He seemed to be worried about his reputation and about the loss of privacy. He wrote on 25 October, 1953, "In the past it never occurred to me that every casual remark of mine would be snatched up and recorded. Otherwise I would have crept further into my shell".[55]

And Einstein wrote Seelig on August 12, 1954, that he was pleased that Seelig's publishing work was successful, but if he was also tempted to address to him many letters, then Einstein warned Seelig that he could not answer too many of them, resulting guilty conscience.[56] And Einstein wrote twice *exactly* the same letter, and therefore one is probably a draft of the letter.[57]

On March 11, 1952, Einstein sent Seelig the following short answer on his Swiss days and special theory of relativity: five to six weeks passed between the conception of the idea of the special theory of relativity and the completion of the relevant publication. Einstein then tells Seelig that before that the arguments and components of the theory had been years in preparation, but without putting forward the ultimate decisive one.[58]

Why did Einstein send Seelig such a short answer? Because Seelig was not a scientist; he was not interested in presenting Einstein's pathway to special relativity; he could not even understand the subtleties and creativity of Einstein. He was interested in the gossip, the anecdotes and personal uninteresting things. And Einstein knew Seelig would not understand what he would tell him. Einstein was interested in creativity, scientific pathways and telling the scientific story of his life – a *scientific biography*. Einstein seemed to be reluctant to cooperate with authors who were unable to write a scientific biography. Seelig like Reiser was a journalist, and they could not even comprehend and capture the richness of the scientific and creative side of Einstein. And for this reason Einstein seemed to have kept the details of his scientific pathway to his theories beyond the reach of such writers.

---

[51] Einstein, Albert, *Mein Weltbild*, 1934, Amsterdam: Querido Verlag.
[52] Einstein, Albert, *Mein Weltbild*, edited by Carl Seelig, 1953, Zürich: Europa Verlag.
[53] ETH Archives: http://www.library.ethz.ch/en/Resources/Digital-collections/Einstein-Online/Princeton-1933-1955.
[54] For instance, On April 20, 1952, Einstein wrote Seelig, "Ich danke Ihnen herzlich für Ihren ausführlichen und lieben Brief. Zuerst die Beantwortung von Teddys Fragen?" Einstein to Seelig, April 20, 1952, ETH-Bibliothek, Zürich - Archive und Nachlässe.
[55] Dukas and Hoffmann, 1979, p. 22; p. 127.
[56] "Lieber Herr Seelig, [...] Es freut mich auch, dass Ihre publizistische Tätigkeit so guten Erfolg hat, wenn sie auch leider viele verlocken wird, Briefe an mich zu richten, die ich nicht beantworten kann und dabei mein Gewissen belasten". Item 39 063, Einstein Archives.
[57] Item 39 064, Einstein Archives (13 August, 1954).
[58] "Zwischen der Konzeption der Idee der speziellen Relativitätstheorie und der Beendigung der betreffenden Publikation sind fünf oder sechs Wochen vergangen. Es würde aber kaum berechtigt sein, dieses als Geburtstag zu bezeichnen, nachdem doch vorher die Argumente und Bausteine Jahrelang vorbereitet worden waren, allerdings ohne die endgiltige Entscheidung vorher zu bringen". Item 39 013, Einstein Archives.

Seelig corresponded with many people who were close to Einstein both before and while corresponding with Einstein, and he wrote the first version of his Einstein documentary biography in 1952: *Albert Einstein Und Die Scweiz*.[59] Seelig soon expanded the book and changed the title to *Albert Einstein; eine dokumentarische Biographie*.[60]

These two books are based on correspondence with as many as twenty or more people who knew Einstein (the recollections of people who knew him from Aarau, the Polytechnic and the patent office), but *not yet* on correspondence with Einstein. There are differences between the two 1952 Seelig versions.[61]

On **25 February 1952**, Einstein referred Seelig to his *non-personal Autobiographical Notes* in the book (Einstein wrote it in English in the letter): "Einstein, Scientist-Philosopher" (Library of Living Philosophers, Northwestern University, Evanston, ILL.)", and told him he would ask permission and examples from the publication.[62]

A month later on **March 6, 1952**, Einstein wrote Michele Besso from Princeton about Seelig's 1952 biography:[63] "I already know that this good Seelig is currently dealing with my childhood. This is justified to some extent, since the rest of my existence is known in detail, which is not the case, specifically, concerning the years spent in Switzerland. This gives a misleading impression, as if, so to speak, my life had begun in Berlin!"[64]

Therefore, Seelig's 1952 biography is fairly accurate as regards to Einstein's childhood (Aarau and maybe early student days in Zürich). As regards to the "years spent in Switzerland" – 1900 onward and Bern years in the Patent Office, these years are indeed unknown, especially Einstein's pathway to his greatest scientific achievements of 1905.

---

[59] Seelig, Carl (1952a), *Albert Einstein Und Die Scweiz*, 1952, Zürich: Europa Verlag.
[60] Seelig Carl (1952b), *Albert Einstein; eine dokumentarische Biographie*, 1952, Zürich: Europa Verlag.
[61] Einstein Archives, Jerusalem.
[62] Item 39 011, Einstein Archives.
[63] Either Seelig, 1952a or Seelig, 1952b.
[64] Einstein to Besso, March 6 1952, letter 182, in Einstein, Albert and Besso, Michele, *Correspondence 1903-1955* translated by Pierre Speziali, 1971, Paris: Hermann.

On **March 30, 1952**, Einstein wrote his friend Maurice Solovine[65]

"Carl Seelig is a good man. But he takes the task that he has undertaken far too seriously, alas, with the result that he bothers everyone. Tell him whatever you think best and pass over whatever you wish in silence. For it is not always good to be presented to the public nude – or rather neuter. Make your decisions but do not communicate them to me, for I do not wish to be mixed up, even indirectly, in this affair. I did of course answer a few positive requests".

Seelig corrected the first editions of his documentary biography in the second edition of 1954.[66] Seelig's 1954 biography adds a few details, after Seelig's later correspondence with Einstein, but not much, and does not add crucial details based on the period in the Patent Office in Bern. In this respect, even Seelig's 1954 biography lacks many details; and one could say in a word that, its originality lies in that it reflects the elder Einstein's anecdotes that he told Seelig during the course of their correspondence, and also the stories that his friends and colleagues were willing to share with Seelig.

Seelig sent this 1954 edition to Einstein, and this was his reaction:[67]

Einstein told Seelig that his secretary Miss Helen Dukas introduced him to his letters. Einstein was impressed of Seelig's sense of humor in his letter. However, it appears he was not as impressed of his biography "Einstein Book" as he was of Seelig's sense of humor in the letter. The book was so trivial or as Einstein put it, somewhat swollen that he even forgot to thank Seelig for sending him the book. It is true Einstein admitted that, the book was the most genuine among the books published about him

---

[65] Einstein, Albert, *Lettres à Maurice Solovine*, 1956, Paris: Gauthier Villars; *Letters to Solovine* (With an Introduction by Maurice Solovine, 1987), 1993, New York: Carol Publishing Group, pp. 130-131.
[66] Seelig Carl, *Albert Einstein; eine dokumentarische Biographie*, 1954, Zürich: Europa Verlag; Seelig Carl, *Albert Einstein: A documentary biography*, Translated to English by Mervyn Savill 1956, London: Staples Press.
[67] Princeton 14. IV.54"

Lieber guter Herr Seelig!

Frl. Dukas machte mich mit Ihrem Briefe Bekannt. Der zeigt mir wieder, was fuer ein feiner Kerl Sie sind, und wie das blinde Schicksal Sie nicht schont. Der Brief zeigt mir aber auch zu meiner Freude, dass Ihr Sinn fuer Humor nicht versiegt ist, […]

Die geaeusserten Bedenken der zweiten, noch etwas angeschwollene Auflage des 'Einstein-Buches' habe ich gar nicht fuer moeglich gehalten. So kommt es, dass ich ganz vergessen habe, Ihnen fuer das uebersandte Exemplar zu danken. Es ist ja wirklich das Aechteste, was ueber mich als Person erschienen ist. Abgesehen von [Philipp] Frank's Buch, das ja aus einem mehr unpersoenlichen Gesichtspunkt geschrieben ist, gibt es ja sonst nichts aus nur einigermassen Zuverlaessiges. Also fangen Sie keine Grillen mehr sondern seien Sie mit sich so zufrieden, wie Sie es verdienen. [...]

Mit herzlichen Gruessen und Wuenschen

Ihr

AE".
Item 39 062, Einstein Archives.

from the personal point of view. Apart from Phillip Frank's book, which is written from a more impersonal point of view, all the rest are only fairly reliable. Einstein told Seelig that he had managed to catch some nuggets, but since he was so satisfied with himself, then he deserves it.

Seelig's 1954 German biography was translated and edited during the translation to English in 1956; and afterwards Seelig sent corrections to the English translation.[68]

## 8. Pais' Einstein

We can guess that, if Einstein had lived long enough to read Abraham Pais' biography from 1982, *'Subtle is the Lord'...The Science and Life of Albert Einstein*, he would have been satisfied with it. A scientific impersonal biography with as few personal details (names, names of places), it tells the life story of Einstein's science, and thus concentrates on the evolution of Einstein's contributions to physics, such as: Entropy and probability, Einstein and the reality of molecules (irreversibility and Brownian motion), Einstein's route to the special theory of relativity and the general theory of relativity, unified field theory, and the quantum theory.[69] The biographical chapters are concise and already include material from the early volumes of the *CPAE*.

John Stachel writes that "Pais presents the most convincing account of Einstein's complex personality I have read, based upon personal contacts as well as documentary material".[70]

He describes it as Pais' biography is a "coherent account of almost everything of scientific significance that Einstein did, along with a great deal of needed historical background information and an eminent physicist's perspective on the significance of Einstein's achievement. This is not done through dry abstracts of the papers but with style and wit, by a man who knew Einstein during his last decade – if not well, perhaps as well as one could know a man like Einstein. […] The book, moreover, is not merely an account of Einstein's scientific work but also includes a biography. Sometimes put into separate sections and sometimes mixed with the account of Einstein's scientific activities, and taking up perhaps a fifth of the book in all, the biographical material constitutes the most accurate account of Einstein's life yet written".[71]

## 9. Einstein's Autobiography

The aging Einstein wrote two Autobiographies: one just before his death in 1955 and the other in 1946, and published in 1949.

---

[68] Item 39 084, Einstein Archives.
[69] Pais, 1982.
[70] Stachel, John, "'Subtle is the Lord'"… The Science and Life of Albert Einstein" by Abraham Pais", *Science* 218, 1982, pp. 989-990; reprinted in Stachel, John, *Einstein from 'B' to 'Z'*, 2002, Washington D.C.: Birkhauser, pp. 551-554; p. 552.
[71] Stachel, 1982, in Stachel 2002, p. 551.

The *Autobiographiche Skizze*. A very short *Skizze* of few pages, which was written in 1955 and starts with his happy life in Aarau, after he had left Munich and the Gymnasium, the German authoritative mentality; Einstein then goes on to speak of his first *Gedanken*-experiment in Aarau, and he dedicates most of the *Skizze* to his close friend to whom he his indebted for invaluable help, Marcel Grossman. The *Skizze* is impersonal and scientific, and very concise, but it is autobiographic in the sense that it is geographic, and it contains names, and thus biographic.[72]

Einstein's *Aautobiographical Notes* from 1949 are different in style and content. Paul Arthur Schilpp, a professor of philosophy persuaded Einstein to write his autobiography for his book dedicated to Einstein. Schilpp had edited a series of books about great living philosophers, and he wanted to edit a book on Einstein as well. Each book was devoted to a single man. It contained his specially written autobiography, followed by a series of essays by authorities evaluating and criticizing his work. These essays were then answered by the philosopher himself. However, Einstein refused to write his Autobiography for the book. Instead he consented to write his *scientific autobiography*. He told Robert Shankland that he even hated to write his *Autobiographical Notes* in German.[73]

Einstein spoke of the *Notes* as his obituary, and when it was done, it was entitled not "Autobiography", but *Autobiographical Notes* (*Autobiographisches* in the original German). It did not begin, as a conventional autobiography might, by saying, "I was born on the 14[th] of March 1879…", and the only names that appear are those of scientists and philosophers. There is a most fleeting reference to his being a Jew: "I came – though the child of entirely irreligious (Jewish) parents – to a deep religiousness […]".[74] The *Notes* "are in no sense geographical. They are essentially placeless. Wherever he went, his ideas went with him, and where he went was here irrelevant".[75]

This exactly suited Einstein. He moved from place to place. He wrote on December 23, 1950, "I have been in America now for seventeen years without having adopted anything of this country's mentality. […] You have never changed your human surroundings and can hardly realize what it is to be an old gipsy. It is not so bad".[76]

---

## 10. How to cross-reference the documentary biographies

It seems that it is preferable to cross-reference the documentary biographies in order to comprehend Einstein's way of thinking, and especially to try to distinguish between Einstein the man and "Einstein the myth", and the many "Einstein anecdotes", told and reproduced again and again as if they were account of actual personal events in Einstein's life.

Seelig, for instance, mentioned Frank. Seelig wrote about Frank's biography: "His [Einstein's] successor [in Prague], Philipp Frank, who during the American emigration wrote an intelligent, amusing but, as regards the Swiss period, an incomplete and not altogether reliable biography of Einstein".[77]

Seelig could dig the above sentence from his memory of reading the letter that Einstein wrote him on February 25, 1952, so that Seelig could turn the last sentence in this letter against Frank.[78] Thus, few of the documentary biographies that were discussed above told exactly the same stories about Einstein, but even when they did, different writers arrived at different conclusions.

On the other hand, Reiser in 1930 had told several of the same stories that appeared later in Seelig's and Frank's biography. They all also drew on some stories from Einstein's sister's biographical sketch. Frank wrote in the introduction to his book, "in so far as pure facts are concerned, I have made partial use of earlier biographies of Einstein. The portrayal of Einstein's personality and of his position in our time, however, derives from my study of the writings of Einstein's friends and enemies, and in large measure from personal conversations with Einstein himself".[79]

Einstein preferred a scientific impersonal biography upon a personal biography, because he considered himself a gipsy scientist, and he had bad memory for personal things. Einstein thought that although the personal aspect should be considered, it must not be made the main thing when telling the story of a scientist. There are few primary sources about Albert Einstein, and we have only fragments of information. Sometimes sources contradict one another, and future sources might be discovered. It is important to stick to primary sources first, before referring to secondary sources reporting about primary sources. However, special care must be taken when considering primary sources and Einstein's opinion of the different biographies that were written before he died in 1955 should be taken into account. It is important to take into consideration Einstein's own remarks and attitude towards the different biographers and others who wrote about his life. We have already seen that Einstein disliked *personal biographies*. In addition, one should be very careful when dealing with secondary sources that rely on other secondary sources.

---

[77] Seelig, 1956, p. 119; Seelig, 1954, p. 143.
[78] "Ich bin völlig damit einverstanden, dass Sie über meine in der Schweiz zu gebrachten entscheidenden Entwicklungsjahre eine kleine Schrift herausgeben, da in der Frank'schen Biographie dieser Lebensabschnitt nur kümmerlich behandelt ist. Andere einigermassen verlässliche Biographien existieren überhaupt nicht". Item 39 011, Einstein Archives.
[79] Frank, 1947/2002, p. xiv.

*I wish to thank Prof. John Stachel from the Center for Einstein Studies in Boston University for sitting with me for many hours discussing special relativity and its history. Almost every day, John came with notes on my draft manuscript, directed me to books in his Einstein collection, and gave me copies of his papers on Einstein, which I read with great interest.*